\documentclass[prl,showpacs,amsmath,amssymb,twocolumn,10pt]{revtex4}
\usepackage[thinlines,thiklines]{easybmat}
\usepackage{amsthm}
\usepackage{enumerate}
\usepackage{color}
\usepackage{dcolumn}
\usepackage{bm}
\usepackage{graphicx}

\newcommand{\comment}[1]{}
\def\>{\rangle}
\def\<{\langle}
\def\Span{\textrm{span}}

\def\Pr{\textrm{Pr}}

\def\va#1{{\texttt{#1}}}
\def\ob#1{#1}
\def\defeq{\stackrel{\textrm{def}}{=}}

\def\Alive{A}
\def\Dead{D}
\def\Female{F}
\def\Male{M}
\def\Treated{T}
\def\Untreated{U}

\begin{document}
\date{December 28, 2011}
\title{\Large {\bf Quantum Simpson's Paradox and High Order Bell-Tsirelson Inequalities}}
\author{Yaoyun Shi}
\email{shiyy@eecs.umich.edu} \affiliation{
Department of Electrical Engineering and Computer Science,
University of Michigan, 2260 Hayward Street, Ann Arbor, MI
48109-2121, USA}

\begin{abstract}
The well-known Simpson's Paradox, or Yule-Simpson Effect, in statistics is often illustrated by the following thought experiment:
A drug may be found in a trial to increase the survival rate for both men and women, but decrease the rate for all the subjects as a whole. 
This paradoxical reversal effect has been found in numerous datasets across many disciplines, and is now included in 
most introductory statistics textbooks.
In the language of the drug trial, the effect is impossible, however, if {\em both} treatment groups' survival rates are higher than {\em both} 
control groups'. Here we show that for quantum probabilities, such a reversal remains possible. In particular,
a ``quantum drug'', so to speak, could be {\em life-saving} for both men and women yet {\em deadly} for the whole population.
We further identify a simple inequality on conditional probabilities that must hold classically but 
is violated by our quantum scenarios, and completely characterize the maximum quantum violation.
As polynomial inequalities on entries of the density operator, our inequalities are of degree $6$.
\end{abstract}

\pacs{03.65.Ud, 03.65.Ta, 03.67.-a, 02.50.-r} \maketitle
\begin{figure*}
\centering
\includegraphics[width=5.5in]{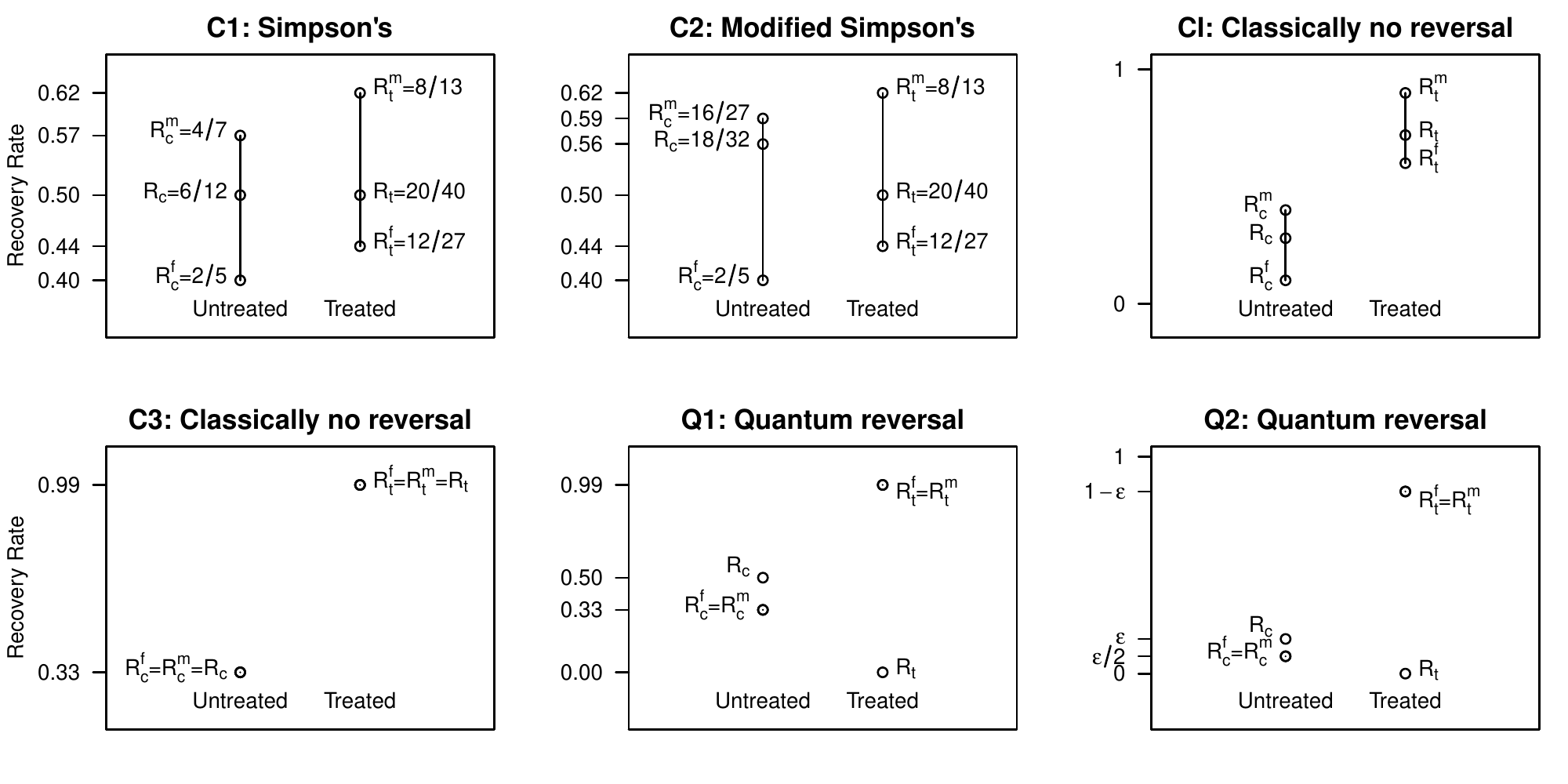}
\caption{{\bf The survival rates in scenarios for the classical and quantum Simpson's Paradox}.
Simpson's original example C1 shows that the positive treatment effect among the males and the females disappears when the two groups are combined.
Each fraction is the number of surviving subjects divided by the total number of subjects in the corresponding group.
Changing the subject numbers in the male treatment group leads to a reversal C2. Q1 and Q2 demonstrate classically impossible quantum reversals,
as in both scenarios, the treatment interval is far higher the control interval,
yet the combined treatment group's rate falls below that for the combined control group. 
The line segments depict rate intervals.
The convexity property fails simultaneously for the treatment and the control groups. 
The classical scenario C3 is to contrast with Q1 as they behave the same on all subgroups
but C3's combined rates do not change due to the convexity property.
Q2 holds for all sufficiently small $\epsilon>0$ with $O(\epsilon^2)$ precision. It
represents a family of maximum quantum reversal, as $S\approx 2-2\epsilon\to2$.
}
\label{figure:qsimpson}
\end{figure*}

Simpson~\cite{Simpson51} used a thought clinical trial of a drug
to illustrate the possibility that an association found in subgroups of a population
may disappear, or even reversed, on the population as a whole.
Denote by $R^f_t$, $R^m_t$, $R_t$, $R^f_c$, $R^m_c$, and $R_c$ the recovery rates of the female, the male, and the combined treatment groups,
respectively. As depicted in Box C1 of Fig.~\ref{figure:qsimpson}, 
$R^f_t > R^f_c$, $R^m_t > R^m_c$, indicating
a beneficial effect on both the females and the males.  However, for the combined control and treatment groups, $R_t=R_c$,
erasing the beneficial effect. With small modifications as shown in C2,  $R_t<R_c$, reversing the beneficial effect.
Numerous real-life datasets from many fields such as epidemiology, social sciences, psychology, and sports,
have been found to exhibit this reversal. The result, called Simpson's Paradox or the Yule-Simpson Effect, among
several other names, is included in most introductory statistics textbooks. 

To allow a generalization to quantum probabilities, 
we discuss Simpson's Paradox in the following model.
Define a {\em (measurement) scenario} $\mathcal{M}\defeq(|\phi\>, \ob{G}, \ob{E}, \ob{R})$,
where $|\phi\>$ is a quantum state, and
$\ob{G}$, $\ob{E}$, $\ob{R}$ are two-outcome measurements on the state space of $|\phi\>$.
For convenience of exposition, we refer to $\ob{G}$, $\ob{E}$, and $\ob{R}$
as the Gender, the Treatment, and the Result measurements, respectively, and name the 
outcomes as { \va{Female}/\va{Male}} ($\Female$/$\Male$),
{\va{Untreated}/\va{Treated}} ($\Untreated$/$\Treated$),
and  {\va{Alive}/\va{Dead}} ($\Alive$/$\Dead$), correspondingly.

Given $\mathcal{M}$, consider two experiments. The first is to measure Gender, followed by Treatment,
and finally Result.
The second is the same, except {\em not} measuring Gender.
The two experiments define the conditional probabilities
$R_t$, $R^f_t$, etc. discussed above. For example, $R_t\defeq\Pr(A|T)$, $R^f_t\defeq\Pr(A|FT)$, etc.

A classical scenario is one where the measurements {\em commute}, i.e., changing the order of applying the measurements does not alter
the outcome statistics. Consequently, $R_t$ can lie anywhere between $R^f_t$ and $R^m_t$ (and likewise for
$R_c$). More precisely, 
\begin{equation}\label{eqn:convex}
R_t=\alpha_f R^f_t + \alpha_m R^f_m,
\end{equation}
where $\alpha_f$ ($\alpha_m$) is the fraction of female (male, respectively) subjects
in the combined treatment group, i.e., $\Pr(F|T)$ (or $\Pr(M|T)$, respectively).
We refer to this relation as the {\em convexity property}.
When the ``treatment interval'' delimited by $R^f_t$ and $R^m_t$ {\em intersects} with the ``control interval'' delimited by $R^f_c$ and $R^m_c$,
a reversal becomes possible (and requires that the gender distributions are different). Conversely, when the two
intervals are {\em disjoint}, i.e., $R^f_t,R^m_t >
R^f_c, R^m_c$, then no reversal is possible. Such is the case depicted in CI, as well as in C3 when $R^f_t=R^m_t=99\%$, and $R^f_c=R^m_c=33\%$.

In sharp contrast, we show that for non-classical scenarios,
a reversal under disjoint rate intervals remains possible.  Q1 in Fig.~\ref{figure:qsimpson} shows that,
with $R^f_t=R^m_t=99\%$ and $R^f_c=R^m_c=33\%$, $R_t$ can be reduced to $0$ and $R_c$ increased to $50\%$,
as opposed to in the classical case remaining at $99\%$ and $33\%$, respectively.
In the more extreme quantum scenario Q2, ``taking the drug'', so to speak,
increases the survival rate
of both gender groups from almost $0$ to almost $1$,
yet reduces the rate for the combined groups to $0$.

Such a reversal is inherently quantum as the convexity property must fail. 
Denote by $d_t=R^f_t+R^m_t-R_t$, $d_c=R^f_c+R^m_c-R_c$ and $S=d_t-d_c$.
For classical scenarios, the convexity property implies that 
$|S|\le 1$. Therefore, the value of
$|S|$ exceeding $1$ in quantum scenarios quantifies intuitively the extent of an inherently quantum reversal. Q1 has $S\approx 7/6$ while
Q2 has $S$ arbitrarily close to $2$. We shall prove that $2$ is precisely the quantum limit, i.e.,  $|S|<2$ for all quantum scenarios.
In particular, a maximum violation of the convexity property cannot occur simultaneously for $R_t$ and $R_c$.

Before we present the details for the construction and the proof, we
sketch below the underlying intuitions.
A fundamental feature of quantum mechanics is that measuring a quantum system may in general {\em change} the state of the system.
Furthermore, two different quantum measurements may be  {\em non-commuting}, a consequence of which is that the state
change incurred by one measurement would alter the outcome statistics of subsequently applying the other.
This is the critical property that underlies our constructions, as well as the Heisenberg Uncertainty Principle,
and other well-known quantum paradoxes such as the EPR Paradox~\cite{EPR35a} and the GHZ Paradox~\cite{GHZ}.

In our construction,  the Gender and the Treatment measurements are non-commuting (though both commute with the Result measurement).
Thus adding the Gender measurement may change dramatically the outcome statistics of the subsequent measurements.
More specifically,  the extremal violation of the convexity property $d_t\to 2$ is a consequence of the following features (illustrated in Fig.~\ref{figure:cube}):
(1) The \va{Treatment} portion of the quantum state has a small amplitude in the \va{Dead} subspace, but a {\em vanishing} amplitude
in the \va{Alive} subspace. This implies $R_t=0$.
(2) After measuring Gender, both eigenstates 
retain a tiny amplitude in the \va{Dead} subspace, but a much larger, even though still small,
amplitude in the \va{Alive} subspace. This
gives $R^f_t=R^m_t\to 1$, and with (1) $d_t\to2$.
Q1 and Q2 are constructed so that $d_t\to2$ and at the same time $d_c$ is made small.
We note that unlike in the violation of spatial Bell inequalities, entanglement is perhaps not relevant 
in our setting.

That $|S|<2$ follows from the necessary tradeoff between $d_t$ and $d_c$.
The tradeoff is most intuitive when considering 
why extremal violation of the convexity property
cannot take place simultaneously on $d_t$ and $d_c$ (in opposite directions).
An extremal violation on $d_t$ require the above two features
(though ``vanishing'' can be replaced by ``very small'').
Consequently, $\Pr(AT)$ and $\Pr(DT)$ are small. 
Similarly, that $d_c\to -2$ implies $\Pr(AU)$ and $\Pr(DU)$ are also small,
contradicting that the sum of those four probabilities is $1$. 

We now discuss the implications of our results and compare them with related works.
Simpson's Paradox is important for revealing the pitfalls of drawing
causal conclusions from partitioned data, and for guiding statistics-based decision making~\cite{Pearl00}. 
Our quantum reversal effect can play a similar role when examining quantum measurement data. With the rapid advances in experimental quantum 
science and engineering, it is anticipated that quantum technologies will be widely used for research and applications in the future. 
Thus there will be more contexts of scientific and practical importance to which our result is directly relevant. 

Our inequalities share the same features of the celebrated Bell Inequalities~\cite{Bel64a, CHSH69a} and Tsirelson Inequalities~\cite{Tsirelson80}
for differentiating classical and quantum probabilities, and for bounding the latter, respectively. 
However, a quantum system is measured only once in Bell-Tsirelson inequalities, while ours involve repeated measurements.
Bell-Tsirelson type inequalities on outcomes of repeated measurements have been discussed by many authors
in the framework introduced by Leggett and Garg~\cite{LeggettG},
and the related framework of ``quantum entanglement in time'' introduced by Brukner et al.~\cite{BruknerTCV}.
Those inequalities are referred to as {\em temporal}, in contrast to the traditional, or {\em spatial}, inequalities.
They bound quantities that are linear in the outcome correlations of repeated measurements. Consequently, they
are on {\em linear} functions of entries of the density operator, just as the spatial Bell-Tsirelson inequalities.
In contrast, our inequalities, in an equivalent form, bound polynomials of higher order (degree $6$). 

Thus, if one interprets broadly the terms Bell and Tsirelson
Inequalities as inequalities bounding functions of the measurement outcome probabilities
for classical, and quantum, respectively, models, our inequalities may be appropriately called high order
Bell-Tsirelson inequalities. We note that quadratic Bell-Tsirelson inequalities were proved by Uffink~\cite{Uffink} in a different context.
We emphasize that the value of such high order inequalities does not lie in the degree per se, but in
that they may, as in our case, provide a different, informative, and intuitive understanding of
quantum effect.  

We speculate that a quantum reversal may also occur in some {\em classical} settings such as human decision making.
For example, in the well-known ``Disjunction Effect''~\cite{TverskyS} experiment,
the human subjects overall preferred Decision $A$ than $B$ on learning {\em either} value of a two-outcome random variable $W$.
Paradoxically, $B$ was preferred instead statistically
when $W$'s value was {\em not} revealed even though the subject knew it was already determined. 
The effect can be interpreted as a consequence of a violation of the convexity property,
similar to our quantum reversal. Note that a ``quantum'' reversal in such a setting does not contradict our earlier conclusion of classical impossibility as the
subject's preference does not have a definite value. This indefinite nature resembles that of the outcome of quantum measurements,
or in the language used by EPR~\cite{EPR35a}, the non-existence of ``elements of reality''.
Indeed, the nascent area of {\em quantum cognition}~\cite{Bruza2009} was partly motivated by the Disjunction Effect
to use quantum probabilities for modeling human cognitive processes.

\def\cubesize{80}
\def\twicecubesize{160}
\def\halfcubesize{40}
\def\dirx{3}
\def\diry{4}
\def\slantx{24}
\def\slanty{32}
\def\lowercx{40}
\def\lowercy{12}
\def\uppercx{40}
\def\uppercy{92}
\def\leftcx{8}
\def\leftcy{48}
\def\rightcx{88}
\def\rightcy{48}
\def\frontcx{35}
\def\frontcy{40}
\def\backcx{60}
\def\backcy{66}

\newcommand{\mycube}[3]{
\def\p{#1}
\def\q{#2}
\begin{picture}(160,160)(0,-20)
\multiput(0,0)(0,\cubesize){2}{\line(1,0){\cubesize}}
\put(\slantx,112){\line(1,0){\cubesize}}
\multiput(\slantx,\slanty)(4,0){20}{\circle*{1}}
\multiput(0,0)(\cubesize,0){2}{\line(0,1){\cubesize}}
\put(104,\slanty){\line(0,1){\cubesize}}
\multiput(\slantx,\slanty)(0,4){20}{\circle*{1}}
\multiput(0,0)(3,4){8}{\circle*{1}}
\put(\cubesize,0){\line(\dirx,\diry){\slantx}}
\multiput(0,\cubesize)(\cubesize,0){2}{\line(\dirx,\diry){\slantx}}
\put(\lowercx,\lowercy){$A$}
\put(\uppercx,\uppercy){$D$}
\put(\leftcx,\leftcy){$T$}
\put(\rightcx,\rightcy){$U$}
\put(\frontcx,\frontcy){$F$}
\put(\backcx,\backcy){$M$}
\multiput(0,0)(\cubesize,0){2}{\circle*{5}}
\multiput(0,\cubesize)(\cubesize,0){2}{\circle*{5}}
\multiput(\slantx,\slanty)(\cubesize,0){2}{\circle*{5}}
\multiput(\slantx,112)(\cubesize,0){2}{\circle*{5}}

\multiput(-18,-3)(\slantx,\slanty){2}{$\frac{\p}{4}$}
\multiput(88,-3)(\slantx, \slanty){2}{$\frac{\p}{4}$}
\put(4,16){$0$}
\put(102,16){$\p$}

\multiput(-20,80)(\slantx, \slanty){2}{$\frac{\q}{4}$}
\multiput(86,80)(\slantx,\slanty){2}{$\frac{\q+2}{4}$}
\put(2,92){$\q$}
\put(98,92){$1$}

\put(\lowercx, -15){#3}
\end{picture}
}

\begin{figure*}
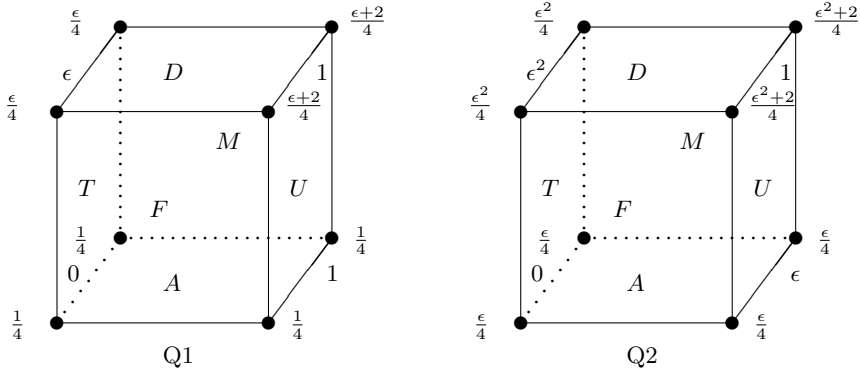

\mycube{1}{\epsilon}{Q1}
\mycube{\epsilon}{\epsilon^2}{Q2}
\caption{\label{figure:cube}
Each of the three axes corresponds to one of the three measurements and each face is annotated with a measurement outcome.
Each face, edge, and vertex represent the resulting state after one, two, or three, respectively, measurements, for the measurement
outcomes determined by the faces incident to it, and annotated with the squared lengths (for those projections in $S$).
For example, Q1's lower-left-closer vertex represents $R^AE^TG^F|\phi\>$ and has a squared length $1/4$.
}
\end{figure*}

{\bf Construction}.
For two quantum states $|\alpha\>$ and $|\beta\>$, denote by
$|(\alpha\pm \beta)\>$ the state $\frac{1}{\sqrt{2}}(|\alpha\>\pm|\beta\>)$. We will use nested
parentheses but may omit the outer-most pair. For example
$|\alpha + (\beta-\gamma)\>\equiv\frac{1}{\sqrt{2}}(|\alpha\> + \frac{1}{\sqrt{2}} (|\beta\>-|\gamma\>))$.

Set $H=V\otimes W$, where $V$ has dimension $4$ with an orthonormal basis
$\{|t\>, |u_0\>, |u_1\>, |u_2\>\}$, and $W$ has dimension $2$ with an orthonormal basis $\{|a\>, |d\>\}$.
The measurement $R$ acts only on $W$, while $G$ and $E$ act on $V$ only.
Denote by $\ob{R}^{A}$ the $\va{Alive}$ eigenspace, or the projection to this eigenspace,
of ${\ob{R}}$, and similarly define $\ob{R}^F$, $\ob{E}^{T}$, etc. 
The measurements are defined through their eigenspaces
\begin{eqnarray*}
\ob{R}^A &\defeq& \Span\{|a\>\},\ \ob{R}^D \defeq \Span\{|d\>\},\\
\ob{E}^T&\defeq&\Span\{|t\>\},\ 
\ob{E}^U \defeq\Span\{|u_i\>:0\le i\le 2\}.
\end{eqnarray*}

To define $G$, we first define the following states:
\begin{eqnarray}
|f_0\>\defeq |(u_0+u_1)+t\>,&& |f_1\>\defeq|u_2\>,\\
|m_0\>\defeq|(u_0+u_1) -t\>,&& |m_1\>\defeq|u_0-u_1\>.
\end{eqnarray}
Now define
\begin{equation*}
\ob{G}^F\defeq\Span\{|f_0\>, |f_1\>\}\ \textrm{and}\ \ob{G}^M\defeq\Span\{|m_0\>, |m_1\>\}.
\end{equation*}
For $p,q\ge0$, define
\begin{eqnarray}
|\phi_a\> &\defeq&\sqrt{p} |u_0+u_1\>,
\textrm{and}\\
|\phi_d\> &\defeq &|(u_0-u_1)+u_2\> + \sqrt{q}|t\>.
\end{eqnarray}

Finally, define (the unnormalized) 
\begin{equation}
|\phi\> \defeq  |\phi_a\>\otimes|a\> +|\phi_d\>\otimes|d\>.
\end{equation}

By direct computation,
\begin{eqnarray}
P(D|T) &=& 1,\\
P(A|U) &=& p/(1+p),\\
P(A|TF)=P(A|TM)&=& p/(p+q),\\
P(D|UF)=P(D|UM)&=& (2+q)/(2+p+q).
\end{eqnarray}

When $p=1$, $q=\epsilon\to0$, we have Q1.
When $q=p^2$ and $p=\epsilon\to0$, we have Q2.
Those two examples are illustrated in Fig.~\ref{figure:cube}.

\def\fp#1#2#3{P\left({#1}/{#2}|#3\right)}

{\bf Bounding the quantum violation}.
We prove here $|S(\mathcal{M})|<2$ for all measurement scenario $\mathcal{M}$. It suffices 
to prove that $S'\defeq S+3 <5$. Without loss of generality we assume that the measurements
are all projective (as otherwise we can replace each POVM by a projective measurement on the extended Hilbert space).
To simplify notation, we define $\ell_{\Dead\Treated}\defeq\|\ob{R}^{\Dead}\ob{E}^{\Treated}|\phi\>\|$,
$\ell_{\Alive\Treated\Female}\defeq \|\ob{R}^{\Alive}\ob{E}^{\Treated}\ob{G}^{\Female}|\phi\>\|$, etc.
and,
\begin{eqnarray}
&&\alpha\defeq \frac{\ell_{AT}}{\ell_{\Dead\Treated}},\ 
 \alpha_F\defeq \frac{\ell_{DTF}}{\ell_{ATF}},\ 
 \alpha_M\defeq \frac{\ell_{DTM}}{\ell_{ATM}},\\
&&\beta\defeq \frac{\ell_{\Dead\Untreated}}{\ell_{\Alive\Untreated}}, \ 
\beta_F\defeq\frac{\ell_{\Alive\Untreated\Female}}{\ell_{\Dead\Untreated\Female}},\ 
\beta_M\defeq\frac{\ell_{\Alive\Untreated\Male}}{\ell_{\Dead\Untreated\Male}}.
\end{eqnarray}
Then
\begin{eqnarray}
S'&=&(1+\alpha^2)^{-1}+(1+\alpha_F^2)^{-1}+(1+\alpha_M^2)^{-1}+\nonumber\\
&& \quad (1+\beta^2)^{-1}+(1+\beta_F^2)^{-1}+(1+\beta_M^2)^{-1}.
\end{eqnarray}
By the triangle inequality,
\begin{equation}\label{eqn:t}
\ell_{DT} \le \ell_{DTF}+\ell_{DTM} = \alpha_F \ell_{ATF} + \alpha_M \ell_{ATM}
\end{equation}
and,
\begin{equation}
\ell_{AU}\le \ell_{AUF} +\ell_{AUM} = \beta_F \ell_{DUF}+\beta_M \ell_{DUM}.
\end{equation}
Decomposing $1=\||\phi\>\|^2$ in two ways, we have
\begin{eqnarray}\label{eqn:master}
&&(1+\alpha^2) \ell_{DT}^2 + (1+\beta^2) \ell_{AU}^2\\
&=& (1+\alpha_F^2)\ell_{ATF}^2+(1+\alpha_M^2)\ell_{ATM}^2+\nonumber\\
&&\quad(1+\beta_F^2)\ell_{DUF}^2+(1+\beta_M^2)\ell_{DUM}^2.
\end{eqnarray}

If $\ell_{DT}=0$ or $\ell_{AU}=0$, $S'\le 5$. It is straightforward to verify that $S'\ne 5$ in either case.
Thus $S'<5$. Consider now that $\ell_{DT}, \ell_{AU}\ne 0$.
Suppose that
\begin{equation}\label{eqn:a}
1+\alpha^2\ge (1+\alpha_F^2)\frac{\ell^2_{ATF}}{\ell^2_{DT}} + (1+\alpha_M^2)\frac{\ell^2_{ATM}}{\ell^2_{DT}}.
\end{equation}
Then
\begin{eqnarray}
&&(1+\alpha^2)^{-1} + (1+\alpha_F^2)^{-1}+(1+\alpha_M^2)^{-1} \nonumber\\
&\le& \left((1+\alpha_F^2)\frac{\ell^2_{ATF}}{\ell^2_{DT}} + (1+\alpha_M^2)\frac{\ell^2_{ATM}}{\ell^2_{DT}}\right)^{-1}\nonumber \\
&&\quad\quad+ (1+\alpha_F^2)^{-1}+(1+\alpha_M^2)^{-1}\\
& \le& 2.
\end{eqnarray}
The second inequality follows by optimizing over all $\frac{\ell_{ATF}}{\ell_{DT}}$ and $\frac{\ell_{ATM}}{\ell_{DT}}$
that satisfy (\ref{eqn:t}). Thus $S'\le 5$. A direct computation shows that equality cannot hold. Thus
$S'<5$.

We need only consider the case when (\ref{eqn:a}) fails. Rearranging Eqn.~(\ref{eqn:master}),
\begin{eqnarray}
&&\ell_{DT}^2\left((1+\alpha^2)-(1+\alpha_F^2)\frac{\ell^2_{ATF}}{\ell^2_{DT}} - (1+ \alpha_M^2)\frac{\ell^2_{ATM}}{\ell^2_{DT}}
\right)\nonumber\\
&& = \ell_{AU}^2\left(-(1+\beta^2)+(1+\beta_F^2)\frac{\ell^2_{DUF}}{\ell^2_{AU}} +\right.\nonumber\\
&&\left.\quad\quad (1+ \beta_M^2)\frac{\ell^2_{DUM}}{\ell^2_{AU}}\right).
\end{eqnarray}
 Thus
 \begin{equation}\label{eqn:u}
 1+\beta^2 > (1+\beta_F^2)\frac{\ell^2_{DUF}}{\ell^2_{AU}} + (1+ \beta_M^2)\frac{\ell^2_{DUM}}{\ell^2_{AU}}.
 \end{equation}
 The proof that $S'<5$ is similar as for the case when (\ref{eqn:a}) holds.
 This completes the proof that $S'<5$ , thus $|S(\mathcal{M})|< 2$, for all measurement scenario $\mathcal{M}$.

I thank Vincent Russo for helpful comments on the writing.
This research was supported in part by the National Basic Research Program of 
China under Awards 2011CBA00300 and 2011CBA00301, and the NSF of the United States 
under Awards 1017335.

\end{document}